\documentclass[%
 reprint,
 amsmath,amssymb,
 aps,
]{revtex4-1}

\usepackage{xcolor}
\usepackage{graphicx}
\usepackage{dcolumn}
\usepackage{bm}
\usepackage{hyperref}

\begin{document}

\title{Effect of local Peregrine soliton emergence on statistics of random waves\\\ in the 1-D focusing Nonlinear Schr\"odinger equation}

\author{Alexey Tikan}
\altaffiliation{University of Lille, UMR 8523-PhLAM-Physique des Lasers
Atomes et Molecules, F-59000 Lille, France}

\email{alexey.tikan@univ-lille.fr}

\date{\today}

\begin{abstract}
The Peregrine soliton is often considered as a prototype of the rogue waves. After recent advances in the semi-classical limit of the 1-D focusing Nonlinear Schr\"odinger (NLS) equation [Bertola, M., Tovbis, A., Commun. Pure Appl. Math. 66, 678–752 (2013)] this conjecture can be seen from another perspective. In the present paper, connecting deterministic and statistical approaches, we numerically demonstrate the effect of the universal local appearance of Peregrine solitons on the evolution of statistical properties of random waves. Evidences of this effect are found in recent experimental studies in the contexts of fiber optics and hydrodynamics. The present approach can serve as a powerful tool for the description of the transient dynamics of random waves and provide new insights into the problem of the rogue waves formation.
\end{abstract}


\maketitle

\section{Introduction}
Ocean waves of extremely high amplitude appear more often than it is predicted by the linear theory~\cite{kharif2008rogue,Onorato:2013Rogue}. Such waves are known as freak or rogue waves (RW). One of the first examples of directly recorded RWs is the New Year wave that crashed onto the Draupner platform in the North Sea in 1995~\cite{Walker:2005The}. The measured wave height is $25.6 \, \mathrm{m}$ (trough to crest), while the significant wave height (four standard deviations of the surface elevation) is approximately $12  \, \mathrm{m}$. Waves of such amplitude unconditionally represent a great danger to mariners. Therefore, understanding the nature of the RWs and predicting their emergence are problems of paramount importance in physics~\cite{kharif2008rogue,Birkholz:2015Predictability,Cousins:2019Predicting}. 

First water tank experiment aiming to demonstrate the emergence of the RWs in well-controlled laboratory conditions is eported in~\cite{Onorato:2004Observation}. The authors studied the nonlinear dynamics of unidirectional waves on the surface of the deep water. In order to mimic the real sea state, a spectrum having a particular asymmetric shape and delta-correlated phases of every Fourier component is used to generate the random initial conditions. This spectrum is found empirically during the measurements provided in the North sea in 1960th known as Joint North Sea Wave Project (JONSWAP)~\cite{hasselmann1973measurements}. According to the central limit theorem, a superposition of a large number of independent Fourier modes leads to the Gaussian distribution for the surface elevation. Therefore, the modulus of the envelope is described by the Rayleigh distribution and the modulus square of the envelope by the exponential distributions ~\cite{ochi:2005ocean}. This wave is also known as partially-coherent. The authors demonstrated that the distribution of the partially-coherent waves deviates from the Gaussian during the propagation in the water tank. This deviation signifies that the probability of extreme events to emerge is increased. Moreover, the rate of the deviation strongly depends on the parameters of the initial conditions and the propagation distance.

\begin{figure}
\center{\includegraphics[width=0.99\linewidth]{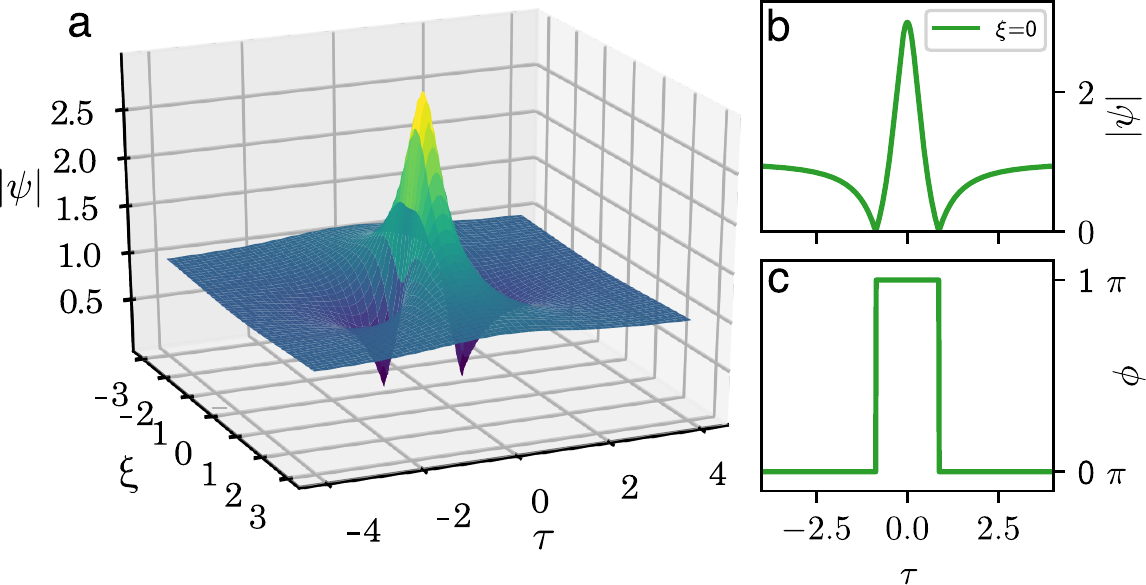}}
\caption{\textbf{Evolution of the Peregrine soliton.} (a) Spatiotemporal diagram of $|\psi|$ plotted according the analytical formula~\cite{Kibler:2010Peregrine}. (b) Cross-section at $\xi$=0. (c) Corresponding phase.}
\label{fig:ps}
\end{figure}

This remarkable observation is in a good agreement with numerical simulations of different hydrodynamic models~\cite{Koussaifi2017Spontaneous}. The leading order model that can be applied to unidirectional surface gravity waves is the 1-D focusing Nonlinear Schr\"odinger (NLS) equation~\cite{Zakharov:1968stab,Onorato:2001Freak,Osborne:2002,Chabchoub:2011Rogue,Onorato:2013Rogue}. NLS is a universal equation that describes the evolution of nonlinear dispersive waves under the assumption of slowly varying envelope. Besides the deep water waves this model governs at the leading order dynamics of electromagnetic waves in a single-mode fiber~\cite{Agrawal:2013} and many other physical systems~\cite{Bao:2007Dynamics}. NLS became a subject of a broad interest after the proof of its integrability with the inverse scattering transform (IST) method~\cite{Zakharov:1972Exact}. It is important to point out that the applicability of envelope models to the random water waves is a subject of ongoing debates, but there are convincing arguments provided in~\cite{Shemer:2010Applicability}.

From integrability of the 1-D focusing NLS equation follows the presence of solitonic solutions. In the context of RWs, an important role plays a family of solitons on the finite background. It is represented by solitons of certain amplitude interacting with a plane wave. Dynamics of such solutions depends on the ratio between the amplitude of the soliton and the level of the plane wave background. There are three well-known members of this family: Akhmediev breather~\cite{akhmediev:1986modulation}, Kuznetsov-Ma soliton~\cite{kuznetsov:1977solitons,Ma:1979} and  Peregrine soliton (PS)~\cite{Peregrine:1983}. Solitons of the finite background are widely considered as prototypes of the RWs~\cite{Osborne:2002,Akhmediev2009Waves,Akhmediev:2009Rogue}. The PS (see Fig.~\ref{fig:ps}) is localized both in space and in time. This property coincides well with the famous characteristic of the RW: it appears out of nowhere and disappears without a trace. Particular importance of the PS in this context is highlighted in~\cite{Shrira:2010What}. Its emergence is usually related to the mechanism of modulation instability of a perturbed plane wave (also known as the Benjamin-Feir instability)~\cite{Dudley:2014Instabilities}.

Presence of the PSs in the spatiotemporal evolution of the partially-coherent wave in the 1-D focusing NLS governed systems is an experimentally verified fact~\cite{Tikan2018Single,Chabchoub:2016Tracking,Cazaubiel:2018Coexistence}. However, the modulation instability cannot be considered as a dominating mechanism in this case. Recent advances in the semi-classical (zero-dispersion) limit of the focusing NLS equation revealed another fundamental mechanism that universally leads to the emergence of the PS. Self-focusing dynamics of a smooth single hump, in the case when nonlinearity significantly dominates dispersion, inevitably leads to a gradient catastrophe. It is proved in~\cite{Bertola:2013} and experimentally verified in~\cite{Tikan:2017} that the structure which appears as a regularization of the gradient catastrophe asymptotically tends the PS. Since the partially-coherent wave can be seen as a set of independent humps at the early stage of the nonlinear propagation, the gradient catastrophe regularization mechanism is expected to play an important role in its evolution. In this article, we numerically demonstrate the influence of the universal local emergence of the PSs on the statistical characteristics of the partially-coherent wave. 

\begin{figure*}
\center{\includegraphics[width=0.9\linewidth]{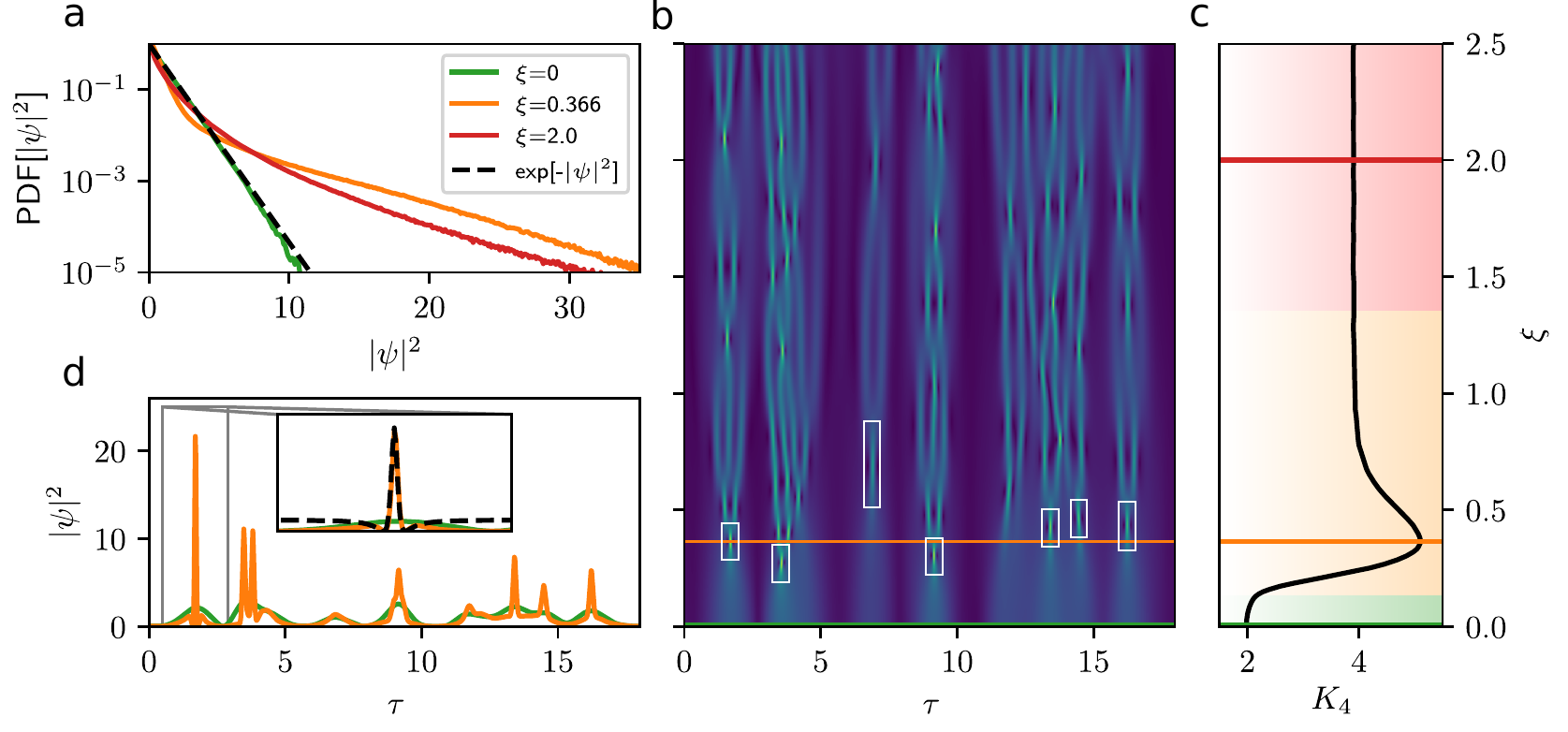}}
\caption{\label{fig:1}\textbf{Numerical simulations of a partially-coherent wave propagation in the focusing NLS system.} (a) Probability density function of $|\psi|^2$ at three different propagation distances $\xi$=0 (green),  0.366 (orange) and 2 (red). The dashed black line corresponds to $\exp(-|\psi|^2)$. (b) Spatiotemporal diagram of $|\psi|$.  White boxes highlight the coherent structures that appear out of initial humps and can be locally fitted with the Peregrine soliton. (c) Evolution of the Kurtosis (black). Green, orange and red lines correspond to $\xi=$0, 0.366, 2. The Kurtosis and the probability density function are computed over 10000 realizations of the partially-coherent wave similar to one depicted in (b). (d) $|\psi|^2$ profile at two values of $\xi$= 0, 0.366 (colors are preserved). Inset plot demonstrates a fit of the hight amplitude event with the exact formula of the Peregrine soliton. In all the simulations the value of $\epsilon$ is 0.2.}
\end{figure*}

\section{Integrable turbulence in NLS equation}

Evolution of random waves in a system governed by an integrable model is considered in the framework of integrable turbulence. The concept of integrable turbulence is introduced by V. E. Zakharov in~\cite{Zakharov:2009turbulence}. In the context of the 1-D focusing NLS equation there are two kinds of initial conditions that are widely studied: a quasi-monochromatic wave with a small random perturbation (condensate)~\cite{Akhmediev2009Waves,Agafontsev2015Integrable,Toenger2015Emergent,Narhi2016Real,kraych:2019statisticalStatistical} and a partially-coherent wave~\cite{Onorato2004Observation,Walczak2015Optical,Chabchoub:2016Tracking,Randoux:2016Nonlinear,Suret2016Single,Koussaifi2017Spontaneous,Tikan2018Single}. 
A transition between these two cases is studied as well using the numerical IST spectra computation~\cite{Akhmediev2016Breather,Soto-Crespo2016Integrable}. 

In the first case, the underlying mechanism is well understood: the monochromatic wave is unstable towards a small perturbation and, therefore, spatiotemporal dynamics is driven by modulation instability. Recent advances in the perturbation approach to this problem allowed to solve exactly the direct and inverse problems for the condensate with a small periodic noise~\cite{Grinevich:2018finite1,Grinevich:2018finite2}. However, it is shown that the evolution of the condensate does not lead to a more frequent emergence of extreme events than it is predicted by the central limit theorem~\cite{Agafontsev2015Integrable}.

Discussion about a mechanism that drives the partially-coherent wave dynamics can be found in~\cite{Suret:2017book}. The authors point out that structures locally similar to the soliton of Peregrine are often found in the numerical simulations and experimental measurements of the partially-coherent wave propagation. Recent water tank experiments confirm the observation~\cite{Cazaubiel:2018Coexistence,Chabchoub:2016Tracking}. 

The definition of the partially-coherent wave used in this article is the following:
\begin{equation}
\psi(\xi,\tau) = \sum\limits_{k} a_k(\xi) \exp^{\frac{2 \pi i}{T}k \tau} \, \text{with } k \in \mathbb{Z},
\label{eq:pcohwave}
\end{equation}
here $a_k(0)=|a_{0k}|e^{i \phi_{0k}}$ is a k$^{th}$ Fourier component with a uniformly distributed random phase $ \phi_{0k} \in [-\pi,\pi]$. This way to introduce the partially-coherent wave guaranties the periodical boundary conditions with the period $T$. 

We write the 1-D focusing Nonlinear Schr\"odinger equation in the following way:
\begin{equation}
i \epsilon \frac{\partial \psi}{\partial \xi} +\frac{ \epsilon^{2}}{2} \frac{\partial^{2} \psi}{\partial \tau^{2}} + |\psi|^{2}\psi = 0 , 
\label{eq:NLSeps}
\end{equation}
where $\epsilon = \sqrt{L_{NL}/L_D}$. In terms of variables adopted in optics it is written as follows: $\epsilon = \sqrt{|\beta_2|/ \gamma P_0 T_0^{2}}$, where $\beta_2$ is the group velocity dispersion coefficient, $\gamma$ - third order nonlinearity coefficient, $P_0 = \frac{1}{T}\int_{0}^{T} <|\psi(\tau, 0)|^2> d\tau$ is the ensemble averaged number of particles and $T_0$ is typical coherence time.

Fig.~\ref{fig:1}b shows a numerically generated spatiotemporal diagram of the partially-coherent wave propagated in the NLS system with $\epsilon$=0.2. Due to the focusing nature of the Eq.~(\ref{eq:NLSeps}), partially-coherent initial conditions can be considered at the early stages of propagation as a set of independent humps, which also follows from the diagram. We also see that if a hump exceeds a certain limit, its dynamics inevitably leads to the formation of a high amplitude coherent structure. It was shown previously that such structures can be \textit{locally} fitted by the PS~\cite{Tikan2018Single,Walczak2015Optical}. Local PSs emerged out of initial humps are highlighted by white rectangles.
Fig. ~\ref{fig:1}d demonstrates cross-sections of the spatiotemporal diagram at $\xi$=0 (green) and 0.366 (orange). As we see, each local hump produces at the first step a single spike during the propagation. The maximum compression point of the spike depends on the initial parameters of each isolated hump. Around the point $\tau$=2 we observe the spike at its maximum compression. The zoomed window in the center shows the fit with the analytical formula of the PS (black dashed line). 
Remarkably, emergence of the PS out of a hump which exceeds certain critical parameter was predicted in~\cite{Shrira:2010What} considering an initial problem with constant boundary conditions.

Statistical properties of this system such as spectrum or probability density function (PDF) evolve together with $\xi$. It is demonstrated in an optical experiment~\cite{Walczak2015Optical} that PDF of $|\psi|^2$ in the case of partially-coherent wave after the propagation in a system governed by the 1-D focusing NLS equation changes from exponential to heavy-tailed (see Fig.~\ref{fig:1}a). It signifies that the probability of the hight amplitude events to appear increases during the nonlinear propagation. This deviation is often characterized by a normalized fourth order moment also known as the Kurtosis. In this paper we use the following definition of the Kurtosis: $K_4=<|\psi|^4>/<|\psi|^2>^2 $, where $<...>$ stands for the ensemble averaging. 

The evolution of the Kurtosis can be divided into three parts (Fig.~\ref{fig:1}c). The first one (green area) corresponds to an early stage of the evolution. In this case, the intensity profile does not experience significant changes, while the phase correlation occurs. The early stage of the evolution of Kurtosis is fully described in the semi-classical limit~\cite{Roberti:2019early}. Deviation of the theory in the vicinity of the upgoing slope was related to the occurrence of the gradient catastrophes in the dynamics of initial humps. 
The second part (orange area) has a well defined \textit{local maximum} (also called overshoot) which is not present in defocusing case~\cite{Onorato:2016origin,Randoux:2016Nonlinear}. At this stage the dynamics of each local hump can still be studied separately from its neighbors. At the distance corresponding to the maximum of $K_4$ the probability to observe the RW is the highest. Moreover, according to~\cite{Onorato:2016origin}, the maximum of $K_4$ corresponds to the maximum of the spectral width.
The third part (red area) corresponds  to the stationary state of the partially-coherent wave dynamics. The description of the stationary state remains an open problem. 

\section{Semi-classical limit of NLS. Universal emergence of the Peregrine soliton}

The emergence of the PS out of smooth rapidly decaying initial conditions is proved in the semi-classical (or zero-dispersion) limit of the 1-D NLS. Zero-dispersion limit implies that the parameter $\epsilon$ in the 1-D NLS equation~(\ref{eq:NLSeps}) tends to zero so the nonlinearity significantly exceeds the dispersion. 

Let's apply so-called Madelung transformation~\cite{Madelung:1927,El:2016Dispersive}:
\begin{equation}
\psi\left ( \xi , \tau\right ) = \sqrt{\rho_m \left ( \xi , \tau\right )} e^{i \phi \left ( \xi, \tau\right ) /\epsilon}, \; u\left ( \xi , \tau\right ) = \phi_{\tau}\left ( \xi, \tau\right ) ,
\label{eq:Change_vars}
\end{equation}
where $ \sqrt{\rho}$ is the wave amplitude and $u$ - the instantaneous frequency.
The 1-D NLS equation~(\ref{eq:NLSeps}) can be expressed as a system of equations by separation of real and imaginary parts:
\begin{eqnarray}
\label{eq:WNLS1}
\rho_{\xi} +(\rho u)_{\tau}=0 \\
\label{eq:WNLS2}
u_{\xi} + uu_{\tau} - \rho_{\tau}   + \frac{\epsilon^{2}}{4} \left[ \frac{\rho_{\tau}^{2}}{2\rho^{2}} - \frac{\rho_{\tau \tau}}{\rho} \right]_{\tau}   = 0.
\end{eqnarray}
In this form, the meaning of the introduced variables can be easily understood. Indeed, the first equation is similar to a continuity equation, with $\rho$ - fluid density and $u$ -  flow velocity field. Together with the second one these equations are analogue of Euler equations for dispersive hydrodynamics but with a negative pressure $p = -\rho^{2}/2$. 

\begin{figure}
\center{\includegraphics[width=0.99\linewidth]{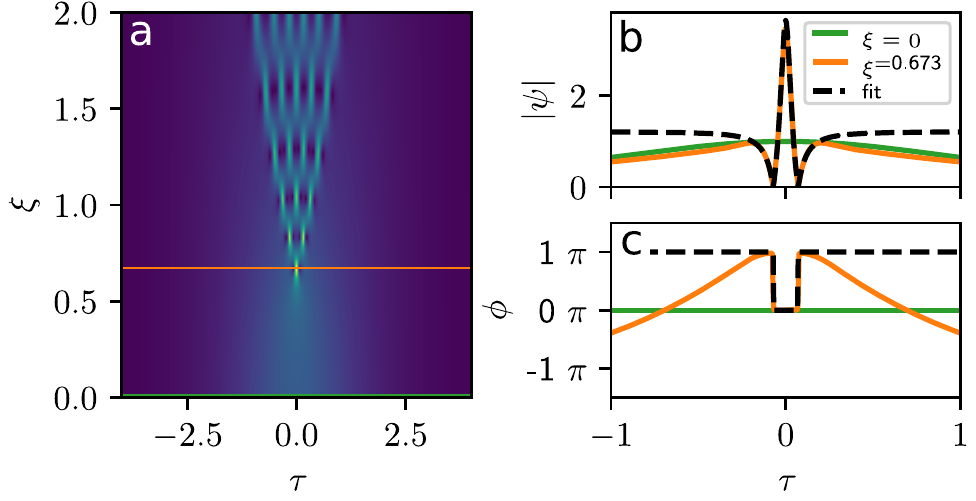}}
\caption{\textbf{Regularization of the gradient catastrophe by local emergence of the Peregrine soliton.} Parameter $\epsilon$ in the simulation is 1/10. (a,b,c) Spatiotemporal diagram, amplitude and phase cross-section at the maximum compression point. $10\mathrm{sech}(\tau)$ function is taken as initial condition. Black dashed line represents a fit of the local PS with the analytical formula (\ref{eq:peregrine}). The maximum compression point in the simulation occurs at the distance $\xi=0.673$ while the distance predicted by expression (\ref{eq:PScompressionPoint}) is $0.6514$.}
\label{fig:intReconstr}
\end{figure}

The last term in Eq.~(\ref{eq:WNLS2}) is proportional to $\epsilon^2$ and therefore it can be neglected at the early stage of the propagation. As it is shown in~\cite{Dubrovin:2009,Bertola:2013}, the propagation of smooth rapidly decaying initial conditions is described by the equations~(\ref{eq:WNLS1},\ref{eq:WNLS2}) until a certain distance where the gradients of $\rho$ or $u$ become (infinitely) large.  This distance is called the point of the \textit{gradient catastrophe}. For the reduced set of equations, the problem is ill-posed and the full system has to be considered. The way to describe solutions in the vicinity of the gradient catastrophe point is to use benefits of the semi-classiacal approximation in the IST method. This is performed by M. Bertola and A. Tovbis in~\cite{Bertola:2013}. They found that the gradient catastrophe is regularized by \textit{universal} appearance of a local coherent structure which asymptotically tends to the PS. Here the term universal is used to underline that Tovbis-Bertola scenario does not depend on the exact shape, chirp or solitonic content of the smooth initial conditions. Similar dynamics is observed in the 1-D NLS equation with linear damping and a Gaussian  driving, as well as a discrete analogue of NLS~\cite{Fotopoulos:201Extreme,Hoffmann:2018Peregrine}.

More explicitly the theory discussed above predicts at the leading order that the point of PS emergence is given by the following expression:
\begin{equation}
\label{eq:PScompressionPoint}
\xi_{m} = \xi_{c} + C\epsilon^{4/5},
\end{equation} 
where $\xi_{m}$ is the maximum compression point, $\xi_{c}=1/2$ is  the point of gradient catastrophe in the case of absence of a chirp, $C = 0.955262458...$ is a universal constant. 
 
The structure that emerges is approximated as:
\begin{equation}
\label{eq:peregrine}
|\psi(\tau, \xi_m)| =  a_0 \left(1-\frac{4}{1+4 a_0^2 (\tau/\epsilon)^2} \right)[1+O(\epsilon^{1/5})],
\end{equation}
which coincides with the formula of PS if $\epsilon \ll 1$.

This result is verified in fiber optics experiments~\cite{Tikan:2017}. Fiber optics experiments demonstrated that this scenario can be observed far beyond the formal range of applicability of the semi-classical limit of the 1-D NLS. Regularization of the gradient catastrophe by emergence of a coherent structure that is \textit{locally} fitted by the PS is observed up to  $\epsilon\approx 0.45$.
 
We demonstrate the PS emergence by the following example. Let's consider initial conditions in the form of exact N solitons solution: $\psi(0,\tau)=\mathrm{sech}(\tau)$ with $\epsilon = 1/N$ (Fig.~\ref{fig:intReconstr}b,c green line $N=10$).
The spatiotemporal diagram of its evolution in the 1-D focusing NLS (Fig.~\ref{fig:intReconstr}a) shows that in the self-focusing dynamics leads to the maximum compression point at $\xi=0.673$. According the semi-classical theory the point of the maximum compression occurs at $\xi=0.6514$.  The cross-sections (Fig.~\ref{fig:intReconstr}b) show that  the coherent structure that emerge is locally similar to PS including the characteristic phase jump of $\pi$ (Fig.~\ref{fig:intReconstr}c). This is in a good agreement with the exact analytical expression (\ref{eq:peregrine}) depicted by the black dashed line in Fig.~\ref{fig:intReconstr}b,c. 

\section{Role of local emergence of Peregrine soliton  in dynamics of partially-waves}

The validity of the semi-classical theory in a wide range of values of $\epsilon$ signifies that the scenario of regularization of the gradient catastrophe by emergence of local PS due to its universality can have a trace in the dynamics of partially-coherent waves. As we demonstrated before, partially-coherent initial conditions could be seen as a set of separated humps at the early stage of the nonlinear propagation. The form of the humps can be complex and irregular, however, the result of Tovbis and Bertola does not depend on the particular shape of initial conditions requiring it only to be smooth. Therefore, we expect that evolution of every individual hump having a value of $\epsilon_{loc}$ below of a certain threshold will lead to the appearance of localized PS.  

Partially-coherent wave has a typical coherence time $T_h\sim 1/\Delta\nu$, where $\Delta\nu$ is a spectral width. Moreover, the distribution of amplitudes of the local humps has a maximum which depends on the average number of particles $P_0$. Therefore, we can conclude that the distribution of values of $\epsilon_{loc}$  locally computed for each hump will also have a certain maximum. 
 Presence of statistically most probable value of locally estimated $\epsilon_{loc}$ implies that there is a propagation distance at which the emergence of the PS is the most probable. 

The probability distribution of PS to appear as a function of $\xi$ can be estimated numerically. We generate the partially-coherent initial conditions in the Fourier space according to the expression~(\ref{eq:pcohwave}). The average number of particles of the partially-coherent wave $P_0$ is set to 1 as well as the typical initial humps duration. In order to estimate the values of $\epsilon_{loc}$, we detect each local hump in the initial conditions. In order to take into account only the humps that will produce the PS and contribute significantly to the statistics we put a threshold $|\psi(\tau, 0)|^2_{th}=2.5 P_0$ found empirically. We exclude double or multi-humps structures requiring the minimal distance between the considered peaks to be more than 2. The value of $\epsilon_{loc}$ is estimated as follows:
\begin{equation*}
\epsilon_{loc} = \epsilon /(T_{loc}\sqrt{P_{loc}}),
\label{eq:eps_loc}
\end{equation*}
where $\epsilon = 0.2$, $T_{loc}$ is the duration of the hump at the half maximum of amplitude and $P_{loc}$ is the square of maximum value of the local hump amplitude. Having the distribution of $\epsilon_{loc}$, we can apply Eq.~(\ref{eq:PScompressionPoint}) and find a distribution of the PS emergence distances using the following normalization:

\begin{equation*}
\xi_{ps} = \xi_m T_{loc}/\sqrt{P_{loc}}
\end{equation*}

\begin{figure}
\center{\includegraphics[width=0.9\linewidth]{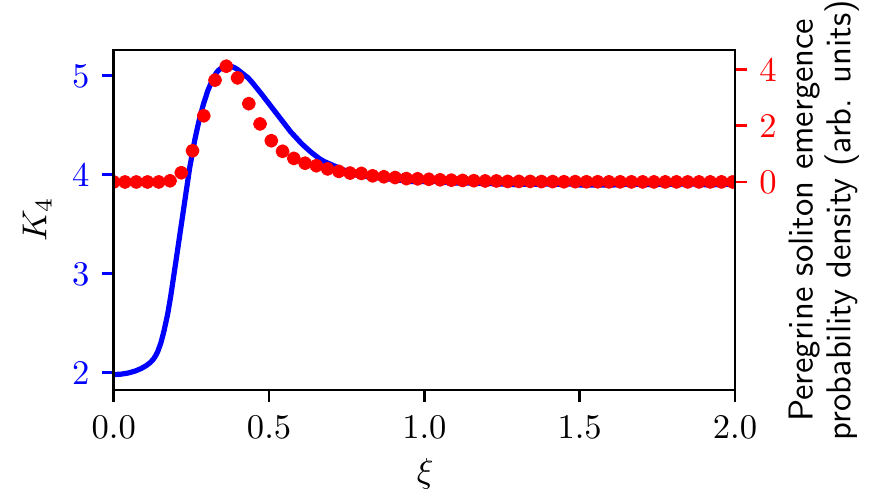}}
\caption{\textbf{Prediction of the maximum of the Kurtosis using the Tovbis-Bertola approach.} Comparison between the probability density of local Peregrine soliton emergence point (red dot, right axis) and the Kurtosis at different propagation distances (blue line, left axis). All parameters coincide with ones in Fig.~\ref{fig:1}}
\label{fig:EpsStat01}
\end{figure}

Fig.~\ref{fig:EpsStat01} shows the comparison between the PS emergence probability density (red dots, right axis) and the Kurtosis (blue line, left axis) (the same as the one depicted in Fig~\ref{fig:1}c). We found a remarkable juxtaposition of the maxima of two curves. The asymmetry of the Kurtosis at its overshoot can be explained by the presence of the next steps of the Tovbis-Bertola scenario in the evolution of the partially-coherent wave. Indeed, the emergence of double-peak coherent structures that follow after the local PS are well-seen in the spatiotemporal diagram Fig.~\ref{fig:1}b.

\begin{figure}
\center{\includegraphics[width=0.99\linewidth]{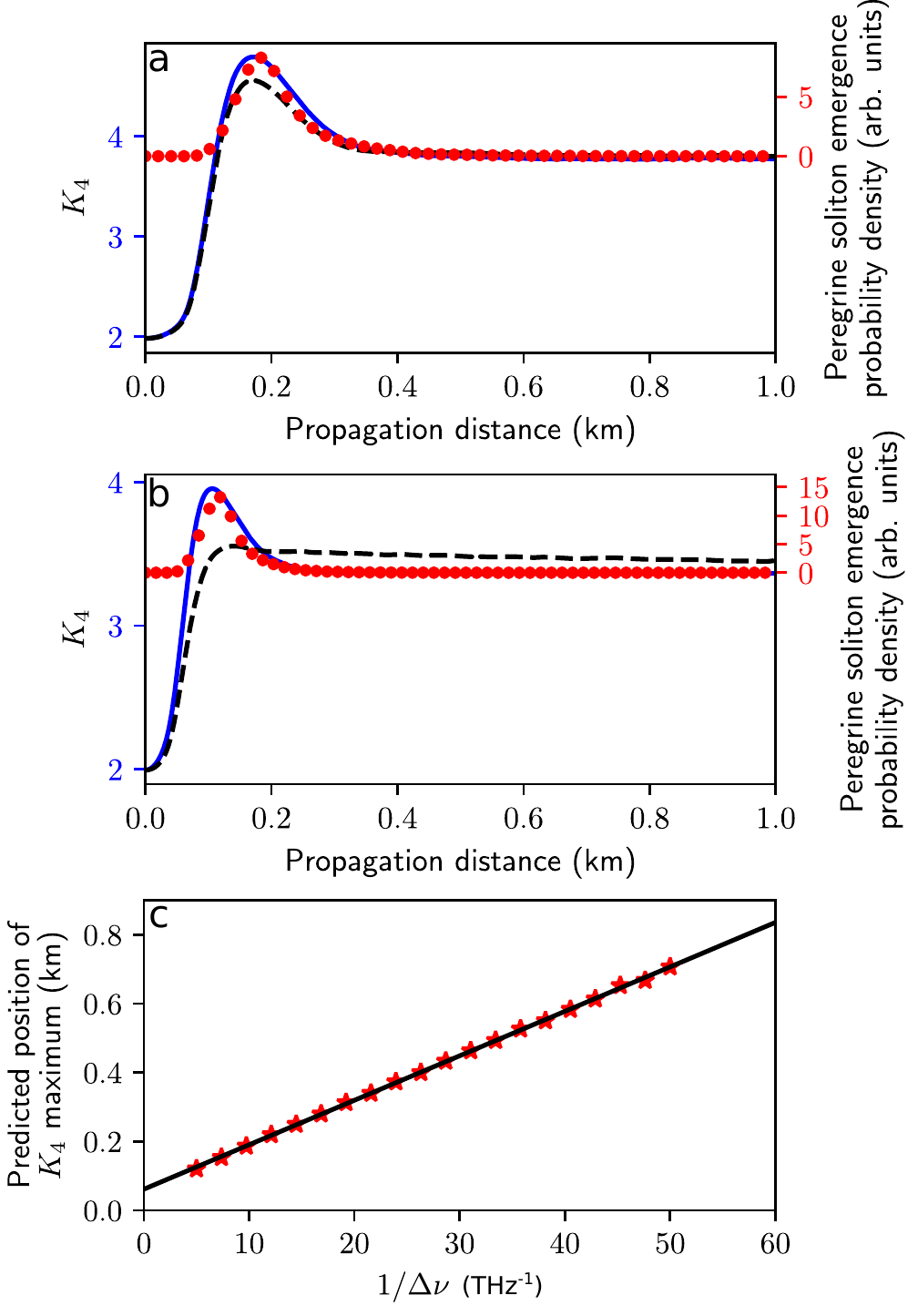}}
\caption{\textbf{Application to the optical fiber experiments}. (a,b) Comparison between the probability density of local Peregrine soliton emergence point (red dot, right axis) and the Kurtosis at different propagation distances (blue line, left axis). Black dashed line corresponds to the Kurtosis of the partially-coherent signal with zero phase. Spectral width of the partially-coherent initial conditions - 0.1 (a) and 0.2 (b)~$\mathrm{THz}$, average power is 2.6~$\mathrm{W}$ for both cases, $\beta_2 = -22 \, \mathrm{ ps^2/km}$, $\gamma=2.4 \, \mathrm{ (W km)^{-1}}$. (c) Prediction for the Kurtosis maximum position for different values of spectral width keeping all other parameters fixed.}
\label{fig:eps_fiber}
\end{figure}

The possibility to predict a position of the maximum of the Kurtosis can be applied to different areas including fiber optics. Following the work~\cite{Tikan2018Single}, we provide estimates of the maximum of the Kurtosis for experimental parameters when propagation of the dynamics is well described by the 1-D NLS equation. We consider two signals of initial spectral widths $0.1 \, \mathrm{THz}$ and $0.2 \, \mathrm{THz}$, average power $2.6 \, \mathrm{W}$. Optical fiber has following characteristics: $\beta_2 = -22 \, \mathrm{ ps^2/km}$, $\gamma=2.4 \, \mathrm{ (W km)^{-1}}$. Fig.~\ref{fig:eps_fiber}a,b show dependence of the Kurtosis on the propagation distance in~$\mathrm{ km}$ (left blue axis, blues curve) superimposed with the dependence of the local PS emergence probability density (right red axis, red dots). The value of nonlinearity in these two cases is less than in the one depicted in Fig.~\ref{fig:EpsStat01}. Therefore, the Kurtosis contains only a trace of the local PS. This also follows from the fact that the width of the overshoot coincides well with the width of the distribution of PS emergence distances.

However, considering parameters far from the semi-classical limit we expect that the role of non-constant phase in partially-coherent wave dynamics is increasing. This follows directly from the expressions~(\ref{eq:Change_vars}). We investigate the influence of the phase by providing similar numerical simulations with exactly the same initial data but setting the phase in the direct space to zero. The evolution of $K_4$ in the zero phase case is shown in~\ref{fig:eps_fiber}a,b by the black dashed lines. For the spectral width of $0.1 \, \mathrm{THz}$ we clearly see a smaller overshoot. Maximum of the overshoot is located at the same distance as in the case of the partially-coherent wave. For the higher spectral width ($0.2 \, \mathrm{THz}$) the maximum of the Kurtosis is significantly shifted forward. Its peak value is close to the one at the stationary state. Remarkably, the Kurtosis in the zero phase case asymptotically tends to the one of the partially-coherent wave at the stationary state.

Assuming the validity of our approach from the spectral width $0.2 \, \mathrm{THz}$ and below, we provide systematic estimates of the position of the maximum of the Kurtosis  for the fixed value of average power $2.6\, \mathrm{W}$. Figure~\ref{fig:eps_fiber}c shows the dependence of the peak value of the Kurtosis as a function of inverse spectral width. In the given range of parameters, the dependence is close to linear.

\section{Conclusion and Discussion}

Thereby, we can conclude that the universal mechanism of gradient catastrophe regularization through the local formation of the Peregrine soliton-like structures plays an important role in the integrable turbulence of the 1-D focusing NLS equation. Its universality leads to the presence of a statistically most probable point of emergence of the Peregrine solitons which is represented by an overshoot in the Kurtosis evolution. 

This process can be considered as a possible explanation of the RW formation problem, of course only at the leading order. Indeed, the experimental works (see for example~\cite{Shemer:2010Applicability}) suggest that the dynamics of surface gravity waves having high values of steepness can significantly deviate from the one predicted by the NSL equation. However, presence of the overshoot in the evolution of the Kurtosis is also found in non-integrable hydrodynamic models such as the Dysthe equation or even the full Euler equations as well as real experimental data~\cite{Onorato:2005Modulational,Shemer:2009experimental,Koussaifi2017Spontaneous}. This fact suggests that the presented way of predicting the maximum of the Kurtosis could be extended to a more general case by taking into the account an influence of higher order terms. The applicability of this approach to non-integrable systems has to be thoughtfully analyzed which is far beyond the scope of this manuscript. 

A part of ideas presented in this article were proposed in the PhD thesis written by the author~\cite{Tikan:2018integrable}.

\begin{acknowledgments}
This work has been partially supported by the Agence Nationale de la Recherche through the LABEX CEMPI project (ANR-11-LABX-0007), the Ministry of Higher Education and Research, Hauts de France council and European Regional Development Fund (ERDF) through the Nord-Pas de Calais Regional Research Council, and the European Regional Development Fund (ERDF) through the Contrat de Projets Etat-Re ́gion (CPER Photonics for Society P4S). The author wants to express his gratitude to Prof. P. Suret and Prof. S. Randoux for guidance, support and possibility to provide independent research. The author is grateful to Prof. G. El for very fruitful discussions. Also, the author thanks G. Michel and A. Cazaubiel for reading the draft and making useful comments.
\end{acknowledgments}

\bibliography{Peregrine}

\end{document}